\documentstyle[12pt]{article}

\global\arraycolsep=1pt
\def \w{\wedge}
\def \bz{\bar z}
\def \bx{\bar x}
\def \bp{\bar \phi}

\begin{document}

\date{KHTP-95-08, \ YITP-95-23, \ \ hep-th/9508012}

\title{Integrable Field Theories derived  
\\ from 4D Self-dual Gravity}

\author{Tatsuya Ueno\thanks{JSPS fellow, No.\,6293.} 
\\ \\ \\
       \it{ Department of Physics, Kyunghee University }\\
       \it{ Tondaemun-gu, Seoul 130-701, Korea}\\
       \it{ and }\\
       \it{ Yukawa Institute for Theoretical Physics }\\
       \it{ Kyoto University, Kyoto 606-01, Japan}
\thanks{Address after January 1996, 
        e-mail: tatsuya@yukawa.kyoto-u.ac.jp } }
\maketitle

\vskip 1.0cm
%%%%%%%ABSTRACT%%%%%%%%%%%%%%%%%%%%%%%%%%%%%%%%%%%%%%%%%%%%%
\abstract{
We reformulate the self-dual Einstein equation as a trio of 
differential form equations for simple two-forms.
Using them, we can quickly show the equivalence of the theory 
and 2D sigma models valued in an infinite-dimensional group, which 
was shown by Park and Husain earlier.
We also derive other field theories including the 2D Higgs bundle 
equation.
This formulation elucidates the relation among those field 
theories.
 }
%%%%%%%%%%%%%%%%%%%%%%%%%%%%%%%%%%%%%%%%%%%%%%%%%%%%%%%%%%%%%%
\thispagestyle{empty}
\newpage

%%%%%%%%%%%%%%%%%%%%%%%%%%%%%%%%%%%%%%%%%%%%%%%%%%%%%%%%%%%%%%%
%%%%%%%%%%%%%%%%%%%%%%%%%%
\section{Introduction}   %
%%%%%%%%%%%%%%%%%%%%%%%%%%
The four-dimensional self-dual Einstein equation (SdE) has been given 
attention for a long time both in physics and mathematics, as well as 
the self-dual Yang-Mills equation. Among a number of works associated 
with the SdE \cite{EGH}, an interesting and important subject is to 
connect it to other (possibly simple) field equations. Well-known 
examples of it are Plebanski's heavenly forms \cite{plebanski}, there 
the SdE is given in terms of one function of space-time coordinates. 
Q-Han Park \cite{park} and Ward \cite{ward} have shown that the SdE is 
derived from several two-dimensional sigma models with the gauge group 
of area preserving diffeomorphisms, SDiff(${\cal N}_2)$. Park also has 
clarified the correspondence between the sigma models and first and 
second heavenly forms. On the other hand, by Ashtekar's canonical 
formulation for general relativity \cite{ashtekar1}, the SdE has been 
reformulated as the Nahm equation \cite{ashtekar2}, and its covariant 
version is given in Ref.\,\cite{mason}. 
Through this formulation, Husain has arrived at one 
of sigma models, that is, the principal chiral model \cite{husain}. 
Also by several reduction methods, other interesting models, e.g. the 
SL$(\infty)$ (affine) Toda equation \cite{boyer}\cite{park}, the KP 
equation \cite{castro} etc., are obtained. 
\par

Although we have various examples connected to the SdE, their relation
is rather unclear since their derivations from the SdE are more or less 
complicated and separated. Such a link of the models, however, should be 
investigated in order to understand the SdE further and in particular to 
develop the quantization of self-dual gravity. 
\par

In this paper, we describe the self-dual Einstein space by a trio of 
differential form equations for simple two-forms and derive several 
integrable theories quickly. This formulation elucidates their relation 
and may indicate the possibility to find further a large class of models 
connected to the SdE.

\vskip 0.4cm
%%%%%%%%%%%%%%%%%%%%%%%%%%%%%%%%%%%%%%
\section{Self-dual Einstein equation}%
%%%%%%%%%%%%%%%%%%%%%%%%%%%%%%%%%%%%%%

We start from the observation that the SdE is expressed as closed-ness 
conditions of basis of the space of anti-self-dual two-forms,
%----------------------------------------------------------------------
\begin{equation}
 d ( e^0 \w e^i - {1 \over 2} \epsilon_{ijk} e^j \w e^k ) = 0 \ ,
 \qquad i = 1,2,3,                                     \label{eq: cls}
\end{equation}
%-----------------------------------------------------------------------
where $e^{0,i} = e^{0,i}_{\mu} dx^{\mu}$ are tetrad one-forms on 
four-manifold. This formulation was employed by Plebanski to reduce 
the SdE to the heavenly forms \cite{plebanski}, there the indices 
$i,j,k$ are replaced with spinor ones $A,B$ by the Pauli matrices 
${(\sigma^i)_A}^B$. 
The equations in (\ref{eq: cls}) appear also in 
Ref.\,\cite{capovilla}\cite{abe}. 
We rewrite (\ref{eq: cls}) by defining a null basis,
%----------------------------------------------------------------------
\begin{eqnarray}
&& {\cal Z} = e^0 + i e^1, \qquad \ {\bar {\cal Z}} = e^0 - i e^1,
\nonumber \\
&& {\cal \chi} = e^2 - i e^3, \qquad \ {\bar {\cal \chi}} 
= e^2 + i e^3 \ .
\end{eqnarray}
%----------------------------------------------------------------------
Then (\ref{eq: cls}) becomes 
%----------------------------------------------------------------------
\begin{equation}
d ( {\cal Z} \w {\cal \chi}) = 0 \ ,
\qquad
d ( {\bar {\cal Z}} \w {\bar {\cal \chi}}) = 0 \ ,
 \qquad 
d ( {\cal Z} \w {\bar {\cal Z}} + {\cal \chi} 
\w {\bar {\cal \chi}}) = 0 \ .                \label{eq: nc}
\end{equation}
%-----------------------------------------------------------------------
\par

As for symmetry, adding to the diffeomorphism invariance, (\ref{eq: nc})
is invariant under the local SL$(2,C)$ (self-dual) transformation, 
in matrix form,
%----------------------------------------------------------------------
\begin{equation}
\left[ \begin{array}{l} {\cal Z}^{'}  \\ 
{\cal \chi}^{'} \end{array} \right] =
\left[ \begin{array}{cc} a \ & b \ \\ c \ & d \ \end{array} \right]
\left[ \begin{array}{l} {\cal Z}  \\ {\cal \chi} \end{array} \right] \ ,
\qquad
\left[ \begin{array}{l} {\bar {\cal Z}}^{'}  \\ {\bar {\cal \chi}}^{'} 
\end{array} \right]
=
\left[ \begin{array}{cc} d \ & -c \  \\ -b \ & a \ \end{array} \right]
\left[ \begin{array}{l} {\bar {\cal Z}}  
\\ {\bar {\cal \chi}} \end{array} 
\right] \ , 
\quad ad-bc=1 \ ,                          \label{eq: sl}
\end{equation}
%----------------------------------------------------------------------- 
and also invariant under the global SL$(2,C)$ (anti-self-dual) transformation 
for the pairs 
$({\bar {\cal Z}}, {\cal \chi})$, $({\cal Z}, {\bar {\cal \chi}})$ 
with the same form as (\ref{eq: sl}).
\par

The two-form in the last equation in (\ref{eq: nc}) is of rank-four, while 
others are of rank-two, but we can re-express the last equation 
by defining two one-forms ${\cal P} = {\cal Z} - {\bar {\cal \chi}}$, 
${\cal Q} = {\cal \chi} + {\bar {\cal Z}}$.
Using them, we obtain the following equations equivalent 
to (\ref{eq: nc}),
%----------------------------------------------------------------------
\begin{equation}
 d ( {\cal Z} \w {\cal \chi}) = 0 \ ,
\qquad 
 d ( {\bar {\cal Z}} \w {\bar {\cal \chi}}) = 0 \ ,
\qquad 
 d ( {\cal P} \w {\cal Q}) = 0 \ .                    \label{eq: smcl}
\end{equation}
%-----------------------------------------------------------------------
Since all two-forms in (\ref{eq: smcl}) are simple and closed, 
they can be written as, on a local coordinate system,
%----------------------------------------------------------------------
\begin{equation}
 {\cal Z} \w {\cal \chi} = dz \w dx \ , 
 \qquad
 {\bar {\cal Z}} \w {\bar {\cal \chi}} = d\bz \w d\bx \ ,
 \qquad
 {\cal P} \w {\cal Q} =  dp \w dq \ ,          \label{eq: sm}
\end{equation}
%----------------------------------------------------------------------- 
where $(z,x,\bz,\bx,p,q)$ are functions. Although we use the notation 
suitable for the real, Euclidean case, we generally consider the complex 
SdE, so the bars in (\ref{eq: sm}) do not mean the complex conjugation in 
usual. {}From the definition of ${\cal P}$ and ${\cal Q}$, two identities 
are obtained,
%----------------------------------------------------------------------
\begin{eqnarray}
 {\cal Z} \w {\cal \chi} \w {\cal P} \w {\cal Q} &&
 = dz \w dx \w dp \w dq = dz \w dx \w d\bz \w d\bx
 = {\cal Z} \w {\cal \chi} \w {\bar {\cal Z}} \w {\bar {\cal \chi}} \ , 
 \nonumber \\
 {\bar {\cal Z}} \w {\bar {\cal \chi}} \w {\cal P} \w {\cal Q} &&
 = d\bz \w d\bx \w dp \w dq = dz \w dx \w d\bz \w d\bx
 = {\cal Z} \w {\cal \chi} \w {\bar {\cal Z}} \w {\bar {\cal \chi}} \ .
                                                      \label{eq: id}
\end{eqnarray}
%----------------------------------------------------------------------- 
(\ref{eq: smcl}), (\ref{eq: sm}) and (\ref{eq: id}) are key equations 
in our formulation. For later use, we define the notation 
$x^a = (z, \bz)$ and $x^k = (p,q)$.
 
\vskip 0.4cm
%%%%%%%%%%%%%%%%%%%%%%%%%%%%%%%%%%%%%%%%%%%%%%%%%%%%%%%%%%%
\section{ Integrable theories derived from the SdE}       %
%%%%%%%%%%%%%%%%%%%%%%%%%%%%%%%%%%%%%%%%%%%%%%%%%%%%%%%%%%%
(a)${\it \, The \ principal \ chiral \ model}$ \\
At first, let us choose $(z,\bz,p,q)$ as four coordinate variables and 
$(x,\bx) = (A_z, A_{\bz})$ as functions of them. 
Then (\ref{eq: id}) reads
%----------------------------------------------------------------------
\begin{equation}
 dz \w dA_z \w dp \w dq = dz \w dA_z \w d\bz \w dA_{\bz} \ ,
 \quad
 d\bz \w dA_{\bz} \w dp \w dq = dz \w dA_z \w d\bz \w dA_{\bz} \ , 
                                                    \label{eq: pcm0}
\end{equation}
%-----------------------------------------------------------------------
from which we can quickly derive the following equations after expanding 
$dA_{z(\bz)} = \partial_a A_{z(\bz)} dx^a + \partial_k A_{z(\bz)}dx^k$ 
and rescaling $A_{z(\bz)}$ by $-2$,
%----------------------------------------------------------------------
\begin{equation}
 \partial_z A_{\bz} + \partial_{\bz} A_z = 0 \ ,
\qquad
F_{z\bz}= \partial_z A_{\bz} -\partial_{\bz} A_z 
+ \{A_z, A_{\bz} \} = 0 \ ,                      \label{eq: pcm}
\end{equation}
%-----------------------------------------------------------------------
where $\{A_z, A_{\bz} \}$ is the Poisson bracket with respect to $(p,q)$.
%% SHUSEI 4 %%%%%%%%%%%%%%%%%%%%%%%%%%%%%%%%%%%%%%%%%%%%%%%%%%%%%%%%%%%%%%%%
Using the potentials $A_{z(\bz)}$, we define generators 
${\cal A}_{z(\bz)} = \{\ \,, A_{z(\bz)}\} 
= \partial_q A_{z(\bz)} \partial_p - \partial_p A_{z(\bz)} \partial_q$ of the 
algebra of area preserving diffeomorphisms sdiff(${\cal N}_2$), which, at each 
space-time point $(z, \bz)$, act on function on the internal surface ${\cal N}_2$ 
parametrized by the coordinates $(p,q)$.
The second equation in (\ref{eq: pcm}) means the generators ${\cal A}_{z(\bz)}$ 
are pure-gauge, ${\cal A}_{z(\bz)} = g^{-1} \partial_{z(\bz)} g$, where $g$ 
is an element of the group SDiff(${\cal N}_2)$. 
Substituting them into the first equation, we obtain
%------------------------------------------------------------------------
\begin{equation}
\partial_{\bz}(g^{-1}\partial_z g) 
+ \partial_z (g^{-1}\partial_{\bz} g) = 0 \ .
\end{equation}
%--------------------------------------------------------------------------
This is precisely the chiral model equation on the $(z, \bz)$ space-time 
with $(p,q)$ treated as coordinates on ${\cal N}_2$.
%% SHUSEI 4 END %%%%%%%%%%%%%%%%%%%%%%%%%%%%%%%%%%%%%%%%%%%%%%%%%%%%%%%%%%%%%
Let us solve $A_{z(\bz)}$ for a single scalar function $\Theta$ by
the first equation in (\ref{eq: pcm}), that is, 
$A_z = 2 \partial_z \Theta$ and $A_{\bz} = - 2 \partial_{\bz} \Theta$. 
Then the second equation becomes 
%----------------------------------------------------------------------
\begin{equation}
  \Theta_{z,\bz} + \Theta_{z,p} \Theta_{\bz, q} 
                 - \Theta_{z,q} \Theta_{\bz, p} = 0 \ ,               
\end{equation}
%----------------------------------------------------------------------- 
where $\Theta_{z, p} = \partial_z \partial_p \Theta$. 
By an adequate gauge condition for the local SL$(2,C)$ symmetry, 
tetrads can take the form,
%----------------------------------------------------------------------
\begin{eqnarray}
{\cal Z} = h^{1 \over 2} dz \ , \qquad &&
{\cal \chi} = - h^{1 \over 2} d\bz 
- h^{-{1 \over 2}} \Theta_{z,k} dx^k \ , 
\nonumber \\ 
{\bar {\cal Z}} = h^{1 \over 2} d\bz \ , \qquad &&
{\bar {\cal \chi}} = \ \, h^{1 \over 2} dz + 
                          h^{-{1 \over 2}} \Theta_{\bz, k}dx^k \ ,
\qquad h = \{\Theta_{\bz}, \Theta_z \} \ , 
\end{eqnarray}
%----------------------------------------------------------------------
and the line element is  
%----------------------------------------------------------------------
\begin{equation}
ds^2  = {\cal Z} \otimes {\bar {\cal Z}} 
      + {\cal \chi} \otimes {\bar {\cal \chi}}
= - \Theta_{a,k} \, dx^a \otimes dx^k 
+ {1 \over \{\Theta_z, \Theta_{\bz} \}} 
\Theta_{z,k} \Theta_{\bz,l} \, dx^k \otimes dx^l \ .               
\end{equation}
%----------------------------------------------------------------------- 
It is obvious that all self-dual metrices are obtained from this sigma 
model. In the case of $\{A_z,A_{\bz}\}=0$, the volume form 
${1 \over 4}({\cal Z} \w {\bar {\cal Z}} 
\w {\cal \chi} \w {\bar {\cal \chi}})$ 
vanishes, which corresponds to a degenerate space-time.

\vskip 0.2cm
%%%%%%%%%%%%%%%%%%%%%%%%%%%%%%%%%%%%%%%%%%%%%%%%%%%%%%%%%%%%%%%%%%%%%%%%
(b)${\it \, The \ topological \ model \ with  \ the \ WZ \ term \ only}$ 
\\
Next we take $(z,x,\bz,\bx)$ as our coordinates and $(p,q) = (B_0, B_1)$ 
as functions of them. After changing the notation 
$(z,x,\bz,\bx)=(z,\bz,p,q)$, (\ref{eq: id}) gives
%----------------------------------------------------------------------
\begin{equation}
 \{B_0, B_1 \}_{(z,\bz)} = 1 \ , \qquad \{B_0, B_1 \} = 1 \ . 
\end{equation}
%----------------------------------------------------------------------- 
The bracket in the first equation is defined with respect to $(z,\bz)$.
These equations are rather unfamiliar, but if we define 
$\partial_k A_{z(\bz)} = \{B_0, B_1 \}_{(z(\bz), x^k)}$, we can 
easily check the integrability $\partial_{[k} \partial_{l]} A_{z(\bz)} 
= 0$ and 
%----------------------------------------------------------------------
\begin{equation}
  \partial_z A_{\bz} - \partial_{\bz} A_z = 0 \ ,
\qquad
  \{A_z, A_{\bz} \} = 1 \ .                           \label{eq: tm2}
\end{equation}
%----------------------------------------------------------------------- 
%%% SHUSEI 5 %%%%%%%%%%%%%%%%%%%%%%%%%%%%%%%%%%%%%%%%%%%%%%%%%%%%%%%%%%% 
Also in this case, generators ${\cal A}_{z(\bz)} = \{\ \,, A_{z(\bz)}\}$ 
of sdiff(${\cal N}_2$) are pure-gauge, 
${\cal A}_{z(\bz)} = g^{-1} \partial_{z(\bz)} g$, 
and from the first equation, we obtain
%%% SHUSEI 5 END %%%%%%%%%%%%%%%%%%%%%%%%%%%%%%%%%%%%%%%%%%%%%%%%%%%%%%%
%------------------------------------------------------------------------
\begin{equation}
\partial_{\bz}(g^{-1}\partial_z g) 
- \partial_z (g^{-1}\partial_{\bz} g) = 0 \ .
\end{equation}
%------------------------------------------------------------------------
This is the topological model derived from 
the lagrangian of the Wess-Zumino term only \cite{park}. 
The potentials $A_{z(\bz)}$ are given in terms of one  
function $\Omega$ by the first equation in (\ref{eq: tm2}). 
Then $A_{z(\bz)}= \partial_{z(\bz)} \Omega$ 
and the second equation becomes
%----------------------------------------------------------------------
\begin{equation}
  \Omega_{z, p} \Omega_{\bz, q} - \Omega_{z, q} \Omega_{\bz, p} = 1 \ ,
                                                        \label{eq: pl}
\end{equation}
%----------------------------------------------------------------------
which is Plebanski's first heavenly form \cite{plebanski}. 
With a gauge condition for the local SL$(2,C)$ symmetry, 
tetrads are given by
%----------------------------------------------------------------------
\begin{equation}
{\cal Z} = dz \ , \quad {\cal \chi} = d\bz \ , \quad 
{\bar {\cal Z}} = \Omega_{z,k} dx^k \ , 
\quad {\bar {\cal \chi}} = \Omega_{\bz,k} dx^k \ , 
\end{equation}
%----------------------------------------------------------------------
and the line element is $ds^2 = \Omega_{a,k} dx^a \otimes dx^k$.

\vskip 0.2cm
%%%%%%%%%%%%%%%%%%%%%%%%%%%%%%%%%%%%%%%%%%%%%%%%%%%%%%%%%%%%%%%%%%%%%%%%
(c) ${\it \, The \ WZW \ model}$ \\
The equation of the Wess-Zumino-Witten model is obtained by dropping the 
term in the right-hand-side of the first equation in (\ref{eq: pcm0}), 
%----------------------------------------------------------------------
\begin{equation}
 dz \w dA_z \w dp \w dq = 0 \ ,    \qquad
 d\bz \w dA_{\bz} \w dp \w dq = dz \w dA_z \w d\bz \w dA_{\bz} \ .
                                                    \label{eq: wzw}
\end{equation}
%----------------------------------------------------------------------
After relabeling $(z,A_z,\bz,A_{\bz},p,q)$ as $(A_z,z,p,q,\bz,A_{\bz})$, 
(\ref{eq: wzw}) becomes 
%----------------------------------------------------------------------
\begin{equation}
  \partial_z A_{\bz} - \partial_{\bz} A_z= 0 \ ,  \qquad
  \{A_z, A_{\bz} \} = 0 \ . 
\end{equation}
%-----------------------------------------------------------------------
 Solving the potentials as in the case (b), we have 
%----------------------------------------------------------------------
\begin{equation}
  \Omega_{z, p} \Omega_{\bz, q} - \Omega_{z, q} \Omega_{\bz, p} = 0 \ .
\end{equation}
%----------------------------------------------------------------------- 
Comparing it with the first heavenly form (\ref{eq: pl}), we see that the 
WZW model describes a fully degenerate space-time \cite{park}.
 
\vskip 0.2cm
%%%%%%%%%%%%%%%%%%%%%%%%%%%%%%%%%%%%%%%%%%%%%%%%%%%%%%%%%%%%%%%%%%%%%%%%
(d) ${\it The \ Higgs \ Bundle \ equation}$ \\
Here, note that it is not necessary to choose all four coordinate 
variables from the set $(z,x,\bz,\bx,p,q)$. Instead, we can regard 
more than two variables in it as functions. This observation enables 
us to obtain a further large class of models connected to the SdE.
\par

Let us examine the observation by taking $(z,\bz)$ as two coordinate
variables and $(x,\bx,p,q) = (\phi,{\bar \phi},B_0,B_1)$ as functions.
Then (\ref{eq: id}) reads
%----------------------------------------------------------------------
\begin{equation}
 dz \w d\phi \w dB_0 \w dB_1 = dz \w d\phi \w d\bz \w d{\bar \phi} 
 \ , \quad
 d\bz \w d{\bar \phi} \w dB_0 \w dB_1 
 = dz \w d\phi \w d\bz \w d{\bar \phi} \ .     \label{eq: hb}
\end{equation}
%----------------------------------------------------------------------- 
As like the case (b), we introduce 
$\partial_k A_{z(\bz)} = \{B_0, B_1 \}_{(z(\bz), x^k)}$, in which 
the coordinates $(p,q)$ are defined as the condition $\{B_0, B_1\}=1$
is satisfied. This condition ensures the integrability
$\partial_{[k} \partial_{l]} A_{z(\bz)} = 0$.
Using $\phi, {\bar \phi}$ and $A_{z(\bz)}$, (\ref{eq: hb}) and another 
identity ${\cal P} \w {\cal Q} \w {\cal P} \w {\cal Q} = 0$ become 
%----------------------------------------------------------------------
\begin{equation}
\partial_{\bz} \phi + \{A_{\bz}, \phi \} = - \{\phi, \bp \} \ ,
\quad
\partial_{z} {\bar \phi} + \{A_z, {\bar \phi} \} = \{\phi, \bp \} \ , 
\quad
F_{z \bz}(A) = 0 \ ,  
\end{equation}
%-----------------------------------------------------------------------  
or, changing $A_z \rightarrow A_z + \phi$ and 
$A_{\bz} \rightarrow A_{\bz} + {\bar \phi}$,
%----------------------------------------------------------------------
\begin{equation}
\partial_{\bz} \phi + \{A_{\bz}, \phi \}  = 0 \ ,
\qquad
\partial_{z} {\bar \phi} + \{A_z, {\bar \phi} \} = 0 \ , 
\qquad
F_{z \bz} = - \{\phi, {\bar \phi}\} \ .                           
\end{equation}
%-----------------------------------------------------------------------
This is the two-dimensional Higgs bundle equation with the group 
SDiff(${\cal N}_2$), which is mentioned in Ref.\,\cite{ward}.
A self-dual Einstein metric is given by a solution of the model
through the tetrads, 
%----------------------------------------------------------------------
\begin{eqnarray}
{\cal Z} = (\partial_p \phi + {g \over h} \partial_p \bp) dz 
          + {1 \over h} \partial_p \bp \partial_{I} \phi \, dx^I \, ,
&& 
{\cal \chi} = (\partial_q \phi + {g \over h} \partial_q \bp) dz 
          + {1 \over h} \partial_q \bp \partial_{I} \phi \, dx^I \, ,
\\
{\bar {\cal Z}} = (\partial_q \bp 
- {{\tilde g} \over h} \partial_q \phi) d\bz 
- {1 \over h} \partial_q \phi \partial_{J} \bp \, dx^J \, ,
&&
{\bar {\cal \chi}} = - (\partial_p \bp 
- {{\tilde g} \over h} \partial_p \phi)d\bz 
+ {1 \over h} \partial_p \phi \partial_{J} \bp \, dx^J \, , 
\nonumber                                                           
\end{eqnarray}
%-----------------------------------------------------------------------
%% SHUSEI 1 %%%%%%%%%%%%%%%%%%%%%%%%%%%%%%%%%%%%%%%%%%%%%%%%%%%%%%%%%%%%%%
where $x^I$ and $x^J$ mean the sets of variables, $x^I=(\bz,p,q)$, 
$x^J=(z,p,q)$, and $h= \{\phi,\bp \}$, $g=\{\phi,A_z \}$ and 
${\tilde g} = \{\bp, A_{\bz}\}$.  
%% SHUSEI 1 END %%%%%%%%%%%%%%%%%%%%%%%%%%%%%%%%%%%%%%%%%%%%%%%%%%%%%%%%%%
The Higgs bundle equation was originally derived from a dimensional 
reduction of the four-dimensional self-dual Yang-Mills 
theory \cite{hitchin}.

\vskip 0.2cm
%%%%%%%%%%%%%%%%%%%%%%%%%%%%%%%%%%%%%%%%%%%%%%%%%%%%%%%%%%%%%%%%%%%%%%%%%
(e) Let us consider another example 
by choosing $(p,q)$ as two coordinates 
and $(z,x,\bz,\bx)= (C_0,C_1,D_0,D_1)$ 
as functions of $(p,q)$ and other 
$(z, \bz)$. Also in this case we define 
$\partial_k A_{z(\bz)} = \{C_0, C_1 \}_{(z(\bz), x^k)}$, 
$\partial_k B_{z(\bz)} = \{D_0, D_1 \}_{(z(\bz), x^k)}$ and impose the 
conditions $\{C_0, C_1\}=1$, $\{D_0, D_1\}=1$ to ensure the 
integrability of $\partial_k A_{z(\bz)}$ and $\partial_k B_{z(\bz)}$. 
Then we obtain  
%----------------------------------------------------------------------
\begin{equation}
F_{z \bz}(A)=  F_{z \bz}(B)= 0 \ , \
\epsilon^{ab} (\partial_a B_b  + \{A_a, B_b \} ) = 0 \ , \  
\epsilon^{ab} (\partial_a A_b  + \{B_a, A_b \} ) = 0 \ . \label{eq: new}
\end{equation}
%-----------------------------------------------------------------------
The first two equations result from  
${\cal Z} \w {\cal \chi} \w {\cal Z} \w {\cal \chi} =
{\bar {\cal Z}} \w {\bar {\cal \chi}} 
\w {\bar {\cal Z}} \w {\bar {\cal \chi}}=0$.
Tetrads in this case are, for example,
%----------------------------------------------------------------------
\begin{eqnarray}
&&{\cal Z} = h^{1 \over 2} dz 
- h^{- {1 \over 2}} \partial_k A_{\bz} dx^k \ , \ \
\ {\cal \chi} = h^{1 \over 2}d\bz 
+ h^{- {1 \over 2}} \partial_k A_z dx^k \ ,
\ h=\{A_z,A_{\bz}\} \ ,
\nonumber \\
&&{\bar {\cal Z}} = - g^{1 \over 2}d\bz 
                  - g^{- {1 \over 2}} \partial_k B_z dx^k \ , 
\ {\bar {\cal \chi}} = g^{1 \over 2}dz 
                  - g^{- {1 \over 2}} \partial_k B_{\bz} dx^k \ , 
\ \, g=\{B_z,B_{\bz}\} \ .                                                        
\end{eqnarray}
%-----------------------------------------------------------------------
In this model, the flat potentials $A_{z(\bz)}$ and $B_{z(\bz)}$ interact 
with each other through the third and fourth equations in (\ref{eq: new}).
 
\vskip 0.4cm
%%%%%%%%%%%%%%%%%%%%%%%%%%%%%%%%%%%%%%
\section{Conclusion}                 %
%%%%%%%%%%%%%%%%%%%%%%%%%%%%%%%%%%%%%%
In this paper, we have shown a formulation 
of the self-dual Einstein space
which leads to low-dimensional field theories quickly and clearly.
Now the relation among those theories is rather clear. 
For example, if we obtain a solution of (a) the principal chiral model, 
then it is straightforward, at least formally, to derive 
the corresponding solution of (b) the topological model or the first 
heavenly form (\ref{eq: pl}); solving the potentials $A_{z(\bz)}$ 
in (\ref{eq: pcm}) for $(p,q)$, relabeling $(z,A_z,\bz,A_{\bz},p,q)$ as  
$(z,\bz,p,q,B_0,B_1)$ and following the step in the case (b). 
Here, for the opposite direction from (b) to (a), 
we give a simple example. 
A solution $\Omega$ of (\ref{eq: pl}) corresponding to the $k=0$ (flat) 
Gibbons-Hawking metric \cite{gibbons}\cite{EGH} is given by
%----------------------------------------------------------------------
\begin{equation}
 \Omega = 2 \epsilon^{-1} \sqrt{\bz q} \sinh{z} \sinh{p} 
        + 2 \epsilon \sqrt{\bz q} \cosh{z} \cosh{p} \ ,   \label{eq: ex}
\end{equation}
%----------------------------------------------------------------------- 
where $\epsilon$ is an arbitrary constant.
We can make $\Omega$ real by setting the complex conjugate condition 
$z^{\ast}=p$, ${\bz}^{\ast}=q$.
Through the relation $\partial_k A_{z(\bz)} 
= \{B_0, B_1 \}_{(z(\bz), x^k)}$, 
the pair $(B_0, B_1)$ can take the form, 
%----------------------------------------------------------------------
\begin{equation}
 B_0 = \epsilon^{-1/2} \sqrt{2\bz} \sinh{z} 
 - \epsilon^{1/2} \sqrt{2q} \cosh{p} \ , \ 
 B_1 =  \epsilon^{1/2} \sqrt{2\bz} \cosh{z} 
 + \epsilon^{-1/2} \sqrt{2q} \sinh{p} \ .          \label{eq: bb}
\end{equation}
%-----------------------------------------------------------------------  
By changing the notation as noted above and solving for $(A_z,A_{\bz})$, 
we obtain a solution of the principal chiral model (\ref{eq: pcm}),
%----------------------------------------------------------------------
\begin{eqnarray}
 A_z  && = - [{{\epsilon^{-1/2} p \sinh{\bz}  
 + \epsilon^{1/2} q \cosh{\bz} }
 \over {\epsilon^{-1} \sinh{z} \sinh{\bz} 
 + \epsilon \cosh{z} \cosh{\bz}}}]^2 \ ,
\nonumber \\ 
 A_{\bz} && = - [{{\epsilon^{1/2} p \cosh{z} 
 - \epsilon^{-1/2} q \sinh{z} }
 \over {\epsilon^{-1} \sinh{z} \sinh{\bz} 
 + \epsilon \cosh{z} \cosh{\bz}}}]^2 \ .
\end{eqnarray}
%-----------------------------------------------------------------------
Also it may be interesting to discuss 
various reduction procedures through 
this formulation. For example, in the case (b) with the above complex 
conjugate condition, suppose that $(A_z, A_{\bz})$ are functions of 
$\bz, q$ and the imaginary part of $(z,p)$ only.
Then changing the imaginary part and $i A_z$, we obtain the 
three-dimensional Laplace equation, which is just the `translational' 
Killing vector reduction by Boyer and Finley \cite{boyer}.
Also for the `rotational' case, the SL$(\infty)$ 
Toda equation is derived 
from (\ref{eq: id}) quickly.
It is intriguing to pursue a further large class 
of models connected to the  
SdE by arranging the functions $(z,\bz,x,\bx,p,q)$ suitably 
and also to investigate the relation among the models.
\par
 
To check the coordinate variables which permit real, Euclidean metrices, 
let us write down the form of ${\cal P} \w {\cal Q}$ explicitly,
%----------------------------------------------------------------------
\begin{equation}
{\cal P} \w {\cal Q} = 
({\cal Z} \w {\cal \chi} + {\bar {\cal Z}} \w {\bar {\cal \chi}}) 
+ ({\cal Z} \w {\bar {\cal Z}} + {\cal \chi} \w {\bar {\cal \chi}}) \ . 
                                             \label{eq: rc}
\end{equation}
%----------------------------------------------------------------------- 
If all tetrads are real, the real, Euclidean case, the first term in 
(\ref{eq: rc}) is real, while the second term pure-imaginary. 
In the case of (a) the principal chiral model, 
suppose $(p,q)$ are real or 
complex conjugate to each other. 
According to it, ${\cal P} \w {\cal Q} = dp \w dq $ becomes real or 
pure-imaginary. 
But then either the first or the second term in (\ref{eq: rc}) vanishes, 
which corresponds to a degenerate space-time, not an interesting case.
Therefore, with such $(p,q)$, the chiral model 
permits only complex metric, 
or signature $(+,+,-,-)$ real metric in which 
%% SHUSEI 2 %%%%%%%%%%%%%%%%%%%%%%%%%%%%%%%%%%%%%%%%%%%%%%%%%%%%%%%%%%
case
%% SHUSEI 2 END %%%%%%%%%%%%%%%%%%%%%%%%%%%%%%%%%%%%%%%%%%%%%%%%%%%%%%  
two of four tetrad one-forms 
may be taken as pure-imaginary. 
In fact $(B_0, B_1)$ in (\ref{eq: bb}), 
which are $(p,q)$ in the case (a), 
are neither real nor complex conjugate to each other. 
\par

The infinite-dimensional group SDiff(${\cal N}_2$) is known to be 
realized as a large N limit of the SU(N) group when the surface 
${\cal N}_2$ is the sphere or the torus \cite{sun}.
Hence an approach to the SdE is to start from the SU(N) principal 
chiral model, there its explicit
%% SHUSEI 3 %%%%%%%%%%%%%%%%%%%%%%%%%%%%%%%%%%%%%%%%%%%%%%%%%%%%%%%%
classical 
%% SHUSEI 3 END %%%%%%%%%%%%%%%%%%%%%%%%%%%%%%%%%%%%%%%%%%%%%%%%%%%%
solutions can be determined by the 
Uhlenbeck's uniton construction \cite{uhlenbeck}.
Adding to the SDiff(${\cal N}_2$), several (hidden) symmetrical 
structures in the SdE have been studied by the sigma model approach 
\cite{park}\cite{husain}\cite{morales}.
Our formulation may be useful to investigate the structure since, 
from our key equations (\ref{eq: sm}),(\ref{eq: id}), we can easily see 
fundamental symmetries of our models, e.g. the conformal invariance on 
the $(z,\bz)$ space and 
SDiff(${\cal N}_2$), before we derive  their field equations explicitly.
 
\newpage 
%%%%%%%%%%%%%ACKNOWLEDGMENT%%%%%%%%%%%%%%%%%%%%%%%%%%%%%%%%%%%%%%%%%%%
I am grateful to Q-Han Park, S. Nam, Ryu Sasaki 
and S. Odake for discussions.
This work is supported by the department of research in Kyunghee 
University and the Japan Society for the Promotion of Science.
%%%%%%%%%%%%%%%%%%%%%%%%%%%%%%%%%%%%%%%%%%%%%%%%%%%%%%%%%%%%%%%%%%%%%%

\vskip 1.0cm
%%%REFERENCES%%%%%%%%%%%%%%%%%%%%%%%%%%%%%%%%%%%%%%%%%%%%%%%%%%%%%%%%%%

%%%%%%%%%%%%%%%%%%%%%%%%%%%%%%%%%%%%%%%%%%%%%%%%%%%%%%%%%%%%%%%%%%%%% 
\end{document}